\documentclass[sor,
]{revtex4-2}

\usepackage{graphicx}
\usepackage{dcolumn}
\usepackage{bm}
\usepackage{physics}

\begin{document}

\preprint{}

\title[]{Single-particle entanglement gives rise to truly nonlocal effects like single-particle steering}

\author{L M {Ar\'evalo Aguilar}}\email{larevalo@fcfm.buap.mx}\affiliation{Facultad de Ciencias F\'isico Matem\'aticas, Benem\'erita Universidad Aut\'onoma de Puebla. 18 Sur y Avenida San Claudio, Col. San Manuel, C.P: 72520, Puebla, Pue., Mexico}

\date{\today}

\begin{abstract}
In 1927, at the Solvay conference, Einstein posed a thought experiment with the primary intention of showing the incompleteness of quantum mechanics; to prove it, he uses the instantaneous nonlocal effects caused by the collapse of the wavefunction of a single particle --the spooky action at a distance--, when a measurement is done. This historical event precede the well-know Einstein-Podolsk-Rosen criticism over the incompleteness of quantum mechanics. Here, by using the Stern-Gerlach experiment (SGE), we demonstrate how the instantaneous nonlocal feature of the collapse of the wavefunction together with the single-particle entanglement  can be used to produce the nonlocal effect of steering, i.e. the single-particle steering. In the steering process Bob gets a quantum state depending on which observable Alice decides to measure.
To accomplish this, we fully exploit the spreading (over large distances) of the entangled wavefunction of the single-particle. In particular, we demonstrate that the nonlocality of the single-particle entangled state allows the particle to "know" which detector Alice is using to steer Bob's state.
Therefore, notwithstanding strong counterarguments, we prove that the single-particle entanglement gives rise to truly nonlocal effects at two far away places. This open the possibility of using the single-particle entanglement for implementing truly nonlocal task. 
\end{abstract}

\keywords{quantum mechanics, steering, single-particle entanglement, quantum nonlocality}
\maketitle

\section{\label{sec:level1}Introduction}

Einstein efforts to cope with the challenge of the conceptual understanding of quantum mechanics has been a source of inspiration for its development; in particular, the fact that the quantum description of physical reality is not compatible with causal locality was critically analized by him. In 1927 Einstein posed a simple thought experiment where  a \textbf{single-particle} experiencing diffraction by a single slit --hence generating an expanding spherical wavefunction-- reaches a screen; as a result, "\textit{the particle must be considered as potentially present with almost constant probability over the whole area of the screen;  however, as soon as it is localized, a peculiar action-at-a-distance must be assumed to take place which prevents the continuously distributed wave in space from producing an effect at two places on the screen}" \cite{jammer,ballentine72,norsen05}. That is to say, the wavefunction must collapse at the point where the particle is detected, hence, different points situated far away must instantaneously be unable to detect the particle. Notice that it was at the Solvay conference where, as far as we know, the phrase "a peculiar action-at-a-distance" was used for the first time by Einstein  but  in the context of a single-particle only, besides it is important to notice that this phrase has not been used in the Einstein-Podolsky-Rosen analysis. Additionally, in a letter written in 1947 to Born, Einstein used the phrase \textit{spooky action at a distance}\cite{born71}. Hence, it would be worth quoting Einstein´s words: "\textit{\ldots  I admit, of course, that there is a considerable amount of validity in the statistical approach\ldots I cannot seriously believe in it because the theory cannot be reconciled with the idea that physics should represent a reality in time and space free from spooky actions at a distance\cite{born71},}" in this sentence Einstein talk about whether or not the wavefunction describes the Born probability for a single particle, i.e. the phrase \textit{spooky actions at a distance} refers also to a single particle. Then, historically, by using this thought experiment Einstein was able to put  the nonlocal effects of quantum mechanics and the collapse of the wavefunction as a distinctness of quantum mechanics  that should be investigated. Some years later from the Solvay coference, in a letter to Scr\"odinger written in 1935, Einstein reframe this thought experiment in terms of boxes, although as a classical analogy only \cite{norsen05}.

\begin{figure}[h!]
   \centering
  \includegraphics[width=110mm]{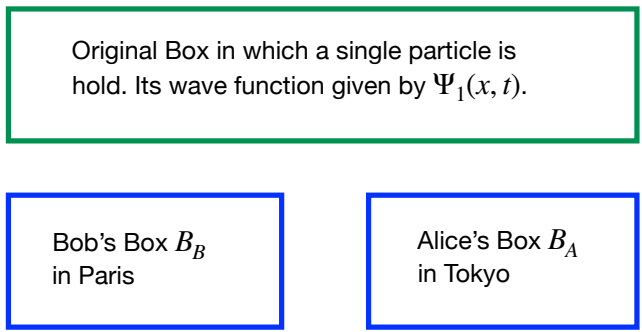}
\caption{\label{paris} A single particle with wavefucntion given by $\Psi_1(x,t)$ is hold in the original Box. After that, the original Box is divide in two Boxes, one of them carried to Paris by Bob and the other one carried to Tokyo by Alice, therefore the wavefunction after the splitting is given by $\Psi_2=\phi_{Paris}(x)+\phi_{Tokyo}(x)$.}
\end{figure}

Meanwhile, de Broglie gave his own version for this though experiment  in 1962 \cite{norsen05}, see also refernce \cite{hardy98}, by using the box thought experiment in which a single particle was situated and where its wave function was given by $\Psi_1(x,t)$; afterwards the box is divided into two boxes, one of them carried to Paris, by Bob, say.  The other one carried by Alice to Tokyo. Hence after this division the wavefunction is given by $\Psi_2=\phi_{Paris}(x)+\phi_{Tokyo}(x)$, see Fig \ref{paris}. Consequently, if Bob opened his box in Paris and found the particle in it, the wave function would collapses to $\Psi_f=\phi_{Paris}(x)$. Besides, de Broglie call attention to the astonishing nonlocal effect of Alice opening her box in Tokyo and finding \textbf{\textit{nothing}} inside, but nevertheless producing the \textbf{\textit{collapse}} of the wavefunction; in this case the wavefunction $\Psi_2=\phi_{Paris}(x)+\phi_{Tokyo}(x)$ also collapses to $\Psi_f=\phi_{Paris}(x)$ in Paris.

The Einstein's boxes resemble the nolocality of single photons, first addressed by Tan et al. \cite{tan91} and further used by Hardy to rule out local hidden variables \cite{hardy94}, see also the work of Peres \cite{peres95}. The nonlocality of single photons has raised great debate which includes their experimental demonstration \cite{hessmo04}, new theoretical and experimental proposal to test it \cite{vedral07, heaney09,jones11,brask13,burch15} and criticism \cite{greenberger,santos,karimi10}. It is worth mentioning that the nonlocal character of the collapse of the  wavefunction was experimentally demonstrated by Fuwa et al. \cite{fuwa15}, see also references \cite{george13,garrisi,guerreiro12}. The single-photon steering was experimentally demonstrated in a detection loophole free scenario by Guerreiro et al. \cite{brunner14b}, the name single-photon steering was stated by N. Brunner \cite{brunner20}.

In the same spirit, the nonlocality phenomenon was also considered in the single-particle entanglement (or intraparticle entanglement), where the entanglement occurs with at least two degree of freedom of a single particle  \cite{azzini20}. It has been addressed, for example, with electrons \cite{dasenbrook16}.  However, the single-particle entanglement faced critical comments over its possible nonlocality properties \cite{karimi10,boyd15, boyd20}; this criticism claims that the entanglement of a single-particle is contextual only, denying nonlocal effects in it. Interestingly, in some of this works the Einstein's phrase over the peculiar action-at-a-distance is only used for the multi-particle entanglement case and ruling out its applicability to the sinlge-particle entanglement situation\cite{boyd15,boyd20,konrad19,forbes19}; however, as it was already stated,  the first time that Einstein himself used that phrase was for  the single-particle case and without considering entanglement. Here, we shown that the single-particle entanglement in the SGE possesses truly nonlocal properties by showing how Alice can steer Bob's states.

On the other hand, before Einstein raised his doubts at the Solvay conference held in 1927, in 1921 Stern proposed an experiment with the aim of testing the Bohr quantization rule of the orbital angular momentum \cite{stern21}. Stern conducted that experiment in 1922 with the help of Gerlach. Meanwhile, in 1925 Uhlenbeck and Goudsmit  proposed the idea of the existence of the internal spin to explain the fine structure phenomena in the espectral emission \cite{weinert95}. However, it was until 1927 that the scientific community began to realize that what the Stern-Gerlach experiment (SGE) really proved was the existence of the internal spin \cite{weinert95}. From then on, the SGE has been a fundamental tool for the development of quantum mechanics; which, as it was explained in references \cite{are2017a,are2017b}, it is an entanglement device. In fact, the quantum attributes of the SGE and a new explanation for how it works and has been given in many papers, see for example reference \cite{paris13}; in particular, in references  \cite{are2017a,are2017b,are19} the Scr\"odinger equation for this experiment has been solved and the violation of local realism was proved in reference \cite{elena2018}. It is well worth mentioning that recent experimental evidence confirms the existence of the superposition of the wavepackets \cite{margalit19,machluf13}. As it was stated in the previous paragraph, in this work we demonstrated  that the quantum properties of the SGE together with the nonlocality of the wavefunction can be used to tailor the steering effect, a process which we can call \textit{single-particle steering}. Steering is a different nonlocal property of quantum mechanics that is stronger that noneparability and weaker that Bell nonlocality \cite{wiseman07,jones07,cavalcanti09,uola20,brunner14,cao,paris20}.


\section{The physics of the Stern-Gerlach experiment}
Here, we analyze how the Stern-Gerlach experiment works in a quantum mechanical way when we send individual atoms one by one. As it was demonstrated in references \cite{are2017a,are2017b} the evolution of the wave function in the Stern-Gerlach experiment is given by:

\begin{eqnarray}
\label{psifinal}
\ket{\psi(t)} =C_0M(x,y)
	\times\Big\{
		e^{\frac{-it\mu_c}{\hbar}(B_{0}+bz)}
		\exp\left[\frac{-1}{4(\sigma_{0}^{2}+i t\hbar/2m)}\left(z+\frac{t^{2}\mu_c b}{2m}\right)^{2}\right] \ket{\uparrow_z} \nonumber\\
	+  e^{\frac{i t\mu_c}{\hbar}(B_{0}+bz)}
		\exp\left[\frac{-1}{4(\sigma_{0}^{2}+i t\hbar/2m)}\left(z-\frac{t^{2}\mu_c b}{2m}\right)^{2}\right]
		\ket{\downarrow_z}\Big\},
\end{eqnarray}
where
\begin{equation}
C_0=e^{\frac{-it^3\mu_c^2b^2}{6m\hbar}}\frac{1}{\sqrt{2}}
		\left[\frac{\sigma_{0}}{(2\pi)^{1/2}}\right]^{3/2}
		\left(\sigma_{0}^{2}+\frac{i \hbar t}{2m}\right)^{-3/2},
\end{equation}
and
\begin{equation}
M(x,y)=e^{-\sigma_0^2k_y^2}
	 e^{\frac{ 4y\sigma_{0}^{2}k_{y}}{4(\sigma_{0}^{2}+ t\hbar/2m)}}
\exp\left[\frac{-(x^{2}+y^{2}-4\sigma_{0}^{4}k_{y}^{2})}{4(\sigma_{0}^{2}+ it\hbar/2m)}\right].
\end{equation}

Now, in order to analize in detail the physics given by this equation: First, notice that what Eq.  (\ref{psifinal}) tell us is that the SGE is an entanglement device, contrary to the usual understanding that considers the SGE as a spin measurement device. That is, there is not any wavefunction collapse in the SGE as it is required by the collapse postulate of quantum mechanics when a measurement is done (Fifth Postulate in reference \cite{cohen-tannoudji2019}), instead what the SGE produces is the entangled state given by Eq.  (\ref{psifinal}). In fact, it was shown that the entangled state given by Eq. (\ref{psifinal}) violates the  Bell's Clauser–Horne–Shimony–Holt  kind inequality \cite{elena2018}.

Second, the wavefunction given in Eq.  (\ref{psifinal}) consists of two Gaussian functions --that are entangled with the spin of the particle--  which are separating one from another in the $z$ axe; at time $t$ the Gaussians are centred, respectively, at position $z=\pm \frac{t^2\mu_c b}{2m}$ , and they are moving at a speed $\frac{\mu_c bt}{2m}$ in opposite directions, which in turn imply a constant acceleration of $\frac{\mu_c b}{2m}$.  Consequently, these Gaussians represent the movement of the external degrees of freedom and correspond to the possible values that these degree of freedom get when measurements are done.  To see a recent demonstration that in quantum mechanics the first Newton law is ruled out see the work of Hofmann \cite{hofmann17a}.


\subsection{Measuring the position observable}
If a screen is put for measuring the possible positions (nowadays a device measure the position by sensing the cloud of electrons that is released when the particle hit it, see Ar\'evalo Aguilar, On the Stern-Gerlach experiment, to be published)  in the $z$ axe, as it was done in the original experiment, two different spots are got. We recall that there exist a connection between physical properties and measurements formulated in the form of a self-adjoint operator \cite{hofmann17}; this connection is realized in the combination of the eigenstate of the operator with a corresponding eigenvalue \cite{hofmann17}. Hence, taking into account that the $z$ position is being measured, then the wavefunction given by Eq. (\ref{psifinal}) collapses (with $50\%$ of probability) towards the eigenstate associated with the eigenvalue of the operator $z$ that was obtained:

\begin{equation}
\label{delta+}
\ket{\psi(t)} =C_0^1M(x,y)  \ket{ Z_+}  \ket{\uparrow_z},
\end{equation}
for the upper spot, where $\ket{Z_+}$ is an eigenket of the operator $z$, and $\braket{z}{Z_+}=\delta\left(z+\frac{t^{2}\mu_c b}{2m}\right)$ is the Dirac delta function, i.e. the wavefunction in $z$ collapses to a well definite position at $z\approx\frac{t^{2}\mu_c b}{2m}$ and $C_0^i$ is normalization coefficient derived from $C_0$, here $i=1$ and below $i=2,3,4,5,6,7,8$. Notice that this spot is still uncertain about the $x$ and $y$ positions and about the momentum $p_z$, in fact the uncertainty in $p_z$ is maximun. The wavefunction given by Eq. (\ref{delta+}) possesses a definite spin $\ket{\uparrow_z}$ and definite position in $z$. However, notice that in this case we are measuring the $z$ position, not the spin, and by detecting the particle at the position eigenstate $\braket{z}{Z_+}=\delta\left(z+\frac{t^{2}\mu_c b}{2m}\right)$  we can infer the spin value and the collapsed  spin state, i.e. $\hbar/2$ and $\ket{\uparrow_z}$ respectively.

At the lower spot the wave function collapses to:

\begin{eqnarray}
\label{delta-}
\ket{\psi(t)} =C_0^2M(x,y)   \ket{ Z_-}    \ket{\downarrow_z},
\end{eqnarray}
where $\ket{Z_-}$ is an eigenket of the operator $z$, and $\braket{z}{Z_-}=\delta\left(z-\frac{t^{2}\mu_c b}{2m}\right)$.
\subsection{Measuring the spin observable}

On the other hand suppose that Alice wants to measure the spin degree of freedom, by using a device that measure the spin (in other words, suppose that a device that measure the spin exists), then the wavefunction given by Eq. (2) collapses (with $50\%$ of probability) towards the eigenstate associated with the value of the obtained eigenvalue (i.e. $+ \hbar/2$) of the operator $\hat{\sigma_z}$:
\begin{eqnarray}
\label{spin+}
\ket{\psi(t)} =C_0^3M(x,y)
                          e^{\frac{-it\mu_c}{\hbar}(B_{0}+bz)}
		e^{\frac{-1}{4(\sigma_{0}^{2}+i t\hbar/2m)}\left(z+\frac{t^{2}\mu_c b}{2m}\right)^{2} }  \ket{\uparrow_z}.
\end{eqnarray}

It is important to highlight that according to the measurement postulate of quantum mechanics what really collapses is the wavefunction associated with the spin degree of freedom (which is what Alice is really measuring), the external degree of freedom in $z$ still is a Gaussian. Also, notice that this wavefunction is still uncertain about the $x$, $y$ and $z$ positions and about the momentum $p_z$. Hence, this spin's measurement is quite different to measure the position as it was done in Eqs. (\ref{delta+}) and (\ref{delta-}). It is worth mentioning that this gives additional evidence to support the statement that the Stern-Gerlach apparatus is not a measurement device, but it is an entanglement device instead.

Similar considerations apply if the eigenvalue $-\hbar/2$ is obtained after the measurement of the spin degree of freedom. In this case, the collapsed wavefunction will be given by:
\begin{eqnarray}
\label{spin-}
\ket{\psi(t)} =C_0^4M(x,y)
                          e^{\frac{it\mu_c}{\hbar}(B_{0}+bz)}
		e^{\frac{-1}{4(\sigma_{0}^{2}+i t\hbar/2m)}\left(z-\frac{t^{2}\mu_c b}{2m}\right)^{2} }  \ket{\downarrow_z}.
\end{eqnarray}




\section{Eintein's boxes with single-particle entanglement}

The de Broglie's version of the Einstein boxes was given in terms of one particle without single-particle entanglement. Consider now the same thought experiment but now  think over that during the splitting process the particle develops single-particle entanglement between the internal and the external degree of freedom. In mathematics: suppose that the initial state of the original box is given by $\ket{\Psi_1(x)}=c_0\ket{\varphi}\left(\ket{\uparrow}+\ket{\downarrow}\right)$ where $\braket{x}{\varphi}=\varphi(x)$ is the $x$ representation of the external degree of freedom, $\ket{\uparrow}$ and $\ket{\downarrow}$ are the internal spin and $c_0$ is a constant. Hence, after the original box's division into two boxes, one in Paris with Bob and the other one in Tokyo with Alice, the wavefunction is given by  $\ket{\Psi_2}=c_0\left(\ket{\varphi_{Paris}}\ket{\uparrow}+\ket{\varphi_{Tokyo}}\ket{\downarrow}\right)$. Consequently, the wavefucntion will collapse depending on what observable Alice decided to measure. For example, if Alice decided to measure the spin observable (and got the value $-\hbar/2$), then the wave function would collapse to $\ket{\varphi_{Tokyo}}\ket{\downarrow}$, where $\braket{x}{\varphi_{Tokyo}}=\varphi_{Tokyo}(x)$. On the other hand, if she decide to measured the position observable (and found the particle at Tokyo), then the wavefunction would collapse to $\ket{\delta_{Tokyo}}\ket{\downarrow}$, where $\braket{x}{\delta_{tokyo}}=\delta(x-x_{Tokyo})$ is the Dirac delta function in $x$.

\begin{figure}[htbp]
\begin{center}
  \includegraphics[width=110mm]{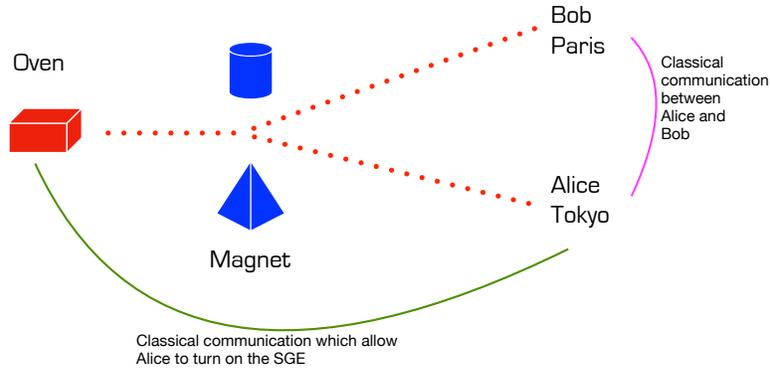}
\caption{\label{tokio} SGE featuring the Einstein's boxes. Where the red box is the oven, the blue box represents the magnet, the red dot represents the fact that there is not classical trajectories, see references \cite{are2017a,are2017b}. Alice could communicate with Bob by using the classical channel in magenta. Moreover, Alice is in full control of the SGE by using the classical channel in green, and she possess the ability to turn it on and to chose between a single or $N$ atoms.}
\end{center}
\end{figure}

Additionally, in the case of the SGE, suppose that just one atom is sent by Alice, and that Alice is in Tokyo and Bob is in Paris. Also, suppose that Alice has automatized the SGE (which is located at suitable place) in such a way that she is able to turn it on (and with the ability to choice whether to send a single atom or $N$ atoms one by one) sending a classical communication, as shown in Fig. \ref{tokio}.

The sep-up of Fig. \ref{tokio} is a scheme of the Einstein's boxes with single-particle entanglement; to perceive this, notice that: 

\begin{enumerate}
\item  If Alice measured the position then (we would recall that Alice is sending a single particle), if her position measuring device registered the particle's position, then the eigenfunction of the particle would collapse to the one given by Eq. (\ref{delta-}), i.e.  $\ \ket{\psi(t)} =C_0^2M(x,y)  \ket{ Z_-}  \ket{\uparrow_z}$.

\item However, if her position measuring device registered nothing, then the eigenfunction of the particle would collapse  --at Bob's location in Paris--  to the one given by Eq. (\ref{delta+}), i.e. $\ket{\psi(t)} =C_0^1M(x,y)  \ket{ Z_+}  \ket{\uparrow_z}$.

\item On the other hand, when Alice decide to measure the spin $\hat{\sigma}_z$, we have:  If her spin measuring device registered the eigenvalue $-\hbar/2$, then the wavefunction would collapses towards Eq. (\ref{spin-}), i.e. $\ket{\psi(t)} =C_0^4M(x,y)
                          e^{\frac{it\mu_c}{\hbar}(B_{0}+bz)}
		e^{\frac{-1}{4(\sigma_{0}^{2}+i t\hbar/2m)}\left(z-\frac{t^{2}\mu_c b}{2m}\right)^{2} }  \ket{\downarrow_z}$.
		
\item  However, if her spin measurement device registered nothing, then the wavefunction would collapses --at Bob's location in Paris-- towards Eq. (\ref{spin+}), i.e. 

$\ket{\psi(t)} =C_0^3M(x,y)
                          e^{\frac{-it\mu_c}{\hbar}(B_{0}+bz)}
		e^{\frac{-1}{4(\sigma_{0}^{2}+i t\hbar/2m)}\left(z+\frac{t^{2}\mu_c b}{2m}\right)^{2} }  \ket{\uparrow_z}$.

\end{enumerate}

\subsection{Still there are more alternatives: Different spin basis.}\label{sssec:num1}

 If Alice decided to measure in a different spin basis, for example $\hat{\sigma}_x$, she could be able to steer a different state. In this case, rewriting Eq. (\ref{psifinal}) in the $\hat{\sigma}_x$ basis, we have:

\begin{equation}
\label{sigmax}
\ket{\psi(t)} =C_0 M(x,y)\Big\{ \left[\braket{z}{\varphi_+}+\braket{z}{\varphi_-}\right]\ket{\uparrow_x} + \left[\braket{z}{\varphi_+}-\braket{z}{\varphi_-}\right]\ket{\downarrow_x} \Big\}
\end{equation}

where
\begin{equation}
\braket{z}{\varphi_+}=e^{\frac{-it\mu_c}{\hbar}(B_{0}+bz)}
		e^{\left[\frac{-1}{4(\sigma_{0}^{2}+i t\hbar/2m)}\left(z+\frac{t^{2}\mu_c b}{2m}\right)^{2}\right] },
\end{equation}

and
\begin{equation}
\braket{z}{\varphi_-}=e^{\frac{it\mu_c}{\hbar}(B_{0}+bz)}
		e^{\left[\frac{-1}{4(\sigma_{0}^{2}+i t\hbar/2m)}\left(z-\frac{t^{2}\mu_c b}{2m}\right)^{2}\right] }.
\end{equation}

Therefore the following alternatives arise:
\begin{enumerate}
\item If her spin $\hat{\sigma}_x$ measuring device registered the eigenvalue $\hbar/2$, then the wavefunction would collapse  towards 
$C_0^5M(x,y) \left[\braket{z}{\varphi_+}+\braket{z}{\varphi_-}\right]\ket{\uparrow_x} $, where $ \left[\braket{z}{\varphi_+}+\braket{z}{\varphi_-}\right]$ is a superposition state of "being" in Paris and Tokyo at the same time.
\item If her spin $\hat{\sigma}_x$ measuring device registered the eigenvalue $-\hbar/2$, then the wavefunction would collapse  towards 
$C_0^6M(x,y) \left[\braket{z}{\varphi_+}-\braket{z}{\varphi_-}\right]\ket{\downarrow_x} $, where $ \left[\braket{z}{\varphi_+}-\braket{z}{\varphi_-}\right]$ is a superposition state of "being" in Paris and Tokyo at the same time.
\end{enumerate}


\subsection{Still there are more alternatives: Momentum}

If Alice decided to measure the momentum $\hat{p}_z$, then the wave function would collapse to the eigenfunction associated with the eigenvalue ($-p_z$) of $\hat{p}_z$ obtained at her momentum measuring device in Tokyo (with the associated spin $\ket{\downarrow_z}$); but if her momentum measuring device detected noting, then the wavefunction would collapse, at a different location --i.e. at Bob's place in Paris-,  towards an eigenfunction of the momentum, i.e. the one associated with the eigenvalue $+p_z$ (with the associated spin $\ket{\uparrow_z}$).



\section{Discussion and Conclusion}

Firstly, there is an unusual situation (a peculiar action-at-a-distance) coming from the above analysis: if Alice measured  $\hat{\sigma}_z$ and detect nothing then the eigenfunction would collapse at Bob's place towards $C_0^3M(x,y) \braket{z}{\varphi_+}  \ket{\uparrow_z}$; however, if Alice measured   $\hat{\sigma}_x$ and detect $\hbar/2$ then the eigenfunction would collapse  towards $C_0^5M(x,y) \left[\braket{z}{\varphi_+}+\braket{z}{\varphi_-}\right]\ket{\uparrow_x} $, but if she detected $-\hbar/2$ the wave function would collapses towards $C_0^6M(x,y) \left[\braket{z}{\varphi_+}-\braket{z}{\varphi_-}\right]\ket{\downarrow_x}$, remember that $ \left[\braket{z}{\varphi_+}\pm\braket{z}{\varphi_-}\right]$ is a superposition state of "being" in Paris and Tokyo at the same time, this fact demonstrates the truly nonlocal feature of the single-particle entanglement: How the particle senses which of the measuring devices Alice is using, i.e. the one that measures   $\hat{\sigma}_z$ or the one that measures $\hat{\sigma}_x$? The answer is that the particle senses which device is in use by the nonlocality of the single-particle entanglement.
		This demonstrates that by measuring the internal degree of freedom, then, the external degree of freedom of the particle might be steered to become a reality in Paris or in Tokyo or even in a superposition state at both places.

Secondly, we can conclude that the spreading of the entangled wavefunction, of the single-particle entangled system, plays a paramount role in the nonlocality features that the system possesses. This open the question to whether such spreading plays a primordial role in the nonlocal properties of some of the multy-particle entangled systems.

Thirdly, the collapse of the wave function depending on what kind of measuring device Alice is using in Tokyo implies some similarity to the Young interferometer, in which  the use of a measuring device near a pinhole plays a primordial role in the loss of the interference pattern. The difference lies in the fact that in the Young interferometer the presence of a measuring device rules out quantum effects whereas in the SGE the presence of a measuring device could be used to produce steering.

Fourthly, what the previous sections prove is that Alice is capable to steer Bob state in Paris by choosing different measurements at Tokio. This is possible due to the spreading over large distances of the entangled wavefunction, i.e. the spreading of the single-particle entangled wavefunction. To produce the steering she ask Bob about the kind of state he wants, then after turning on the SGE by using a classical control protocol (capable of sending a single atom or $N$ atoms, as she wish), she could be able to steer Bob quantum states in such a way that this process rules out the existence of local hidden state models \cite{wiseman07} (using some Bob's protocol if he does not trust Alice). In concrete terms, by the use of $N$ atoms Alice, by means of measuring the position observable, she will obtain $N/2$ times the eigenvalue $-z$ at Tokio, whereas Bob is going to obtain $N/2$ times the eigenvalue $+z$ at Paris. On the other hand, by the use of $N$ atoms Alice, by means of measuring the spin $\hat{\sigma}_z$, she will obtain $N/2$ times the eigenvalue $-\hbar/2$ at Tokio, whereas Bob is going to obtain $N/2$ times the eigenvalue $+\hbar/2$ at Paris. Complex correlations could arise by measuring $\hat{p}_z$ or by measuring $\hat{\sigma}_x$.

Finally, these considerations offers strong support to the fact that single-particle entanglement together with the nonlocal features of its wavefunction can be used to produce nonlocal steering at two different far a way places, superseding contextuality.

\section{Note added in proof}

After completion of the manuscript, we become aware of four papers, the first one is an experimental demonstration of the single-photon steering effect without detection loopholes carried out by Guerreiro et al. \cite{brunner14b}, the second one is an experimental demonstration of the Einstein's thought experimento of 1927 at the Solvay conference \cite{guerreiro12}, the third is a recent experiment testing the collapsing time of a single-photon state \cite{garrisi}. The previous three papers use photons as their test system and we place its citation in the appropriate place in this document. The fourth paper is a theoretical proposal to certify steering for single systems but considering measurements at the same location, i.e. without considering spacelike separation \cite{li15}. These four papers give strong support to our argument. We thanks N. Brunner for letting us know reference \cite{brunner14b}.


\begin{acknowledgments}
We thank Alba Julita Chiyopa Robledo for the proffreading of the manuscript. There was not any support of any scientific agency.
\end{acknowledgments}

\bibliography{plain}

\begin{thebibliography}{50}
\bibitem{jammer} Max Jammer The Philosophy of Quantum Merchanics:
The Interpretations of QM in historical perspective.
John Wiley and Sons (1974).

\bibitem{ballentine72} L. E. Ballentine, Einstein's interpretation of quantum mechanics, Am. J. Phys. \textbf{40}, 1763 (1972).

\bibitem{norsen05} T. Norsen, Einstein's boxes, Am. J. Phys. \textbf{73}, 164 (2005).

\bibitem{born71} M. Born, The Born Einstein Letters, pages 157-159 (Macmillan, London (1971)).

\bibitem{hardy98} L. Hardy, Spooky action at a distance in quantum mechanics, Contemp. Phys. \textbf{39}, 419 (1998).

\bibitem{tan91} S. M. Tan, D. F. Walls, and M. J. Collet, Nonlocality of a single photon, Phys. Rev. Lett. \textbf{66}, 252 (1991).

\bibitem{hardy94} L. Hardy, Nonlocality of a single photon revisited, Phys. Rev. Lett. \textbf{73}, 2279 (1994).

\bibitem{peres95} A. Peres, Nonlocal effects in Fock space, Phys. Rev. Lett. \textbf{74}, 4571 (1995).

\bibitem{hessmo04} B. Hessmo, P. Usachev, H. Heydari, and G. Bj\"ork, Experimental demonstration of single photon nonlocality, Phys. Rev. Letters \textbf{92}, 180401 (2005).

\bibitem{vedral07} J. Dunningham, and V. Vedral, Nonlocality of a single particle, Phys. Rev. Lett. \textbf{99}, 180404 (2007).

\bibitem{heaney09} L. Heaney, and J. Anders, Bell-inequality test for spatial-mode entanglement of a single massive particle, Phys. Rev. A \textbf{80}, 032104 (2009).

\bibitem{jones11}  S. J. Joneas and H. M. Wiseman, Nonlocality of a single photon: Paths to an Einstein-Podolsky-Rosen-steering experiment, Phys. Rev. A \textbf{84}, 012110 (2011).

\bibitem{brask13}  J. B. Brask, R. Chaves, and N. Brunner, Testing nonlocality of a single photon without a shared reference frame, Phys. Rev. A \textbf{88}, 012111 (2013).

\bibitem{burch15}  E. T. Burch, C. Henelsmith, W. Larson, and M. Beck, Quantum-state tomography of a single photon entangled state, Phys. Rev. A \textbf{92}, 032328 (2015).

\bibitem{greenberger}  D. M. Greenberger, M. A. Horne, and A. Zeilinger, Phys. Rev. Lett. \textbf{75}, 2064 (1995).

\bibitem{santos} E. Santos, Comment on ‘‘Nonlocality of a single photon’’, Phys. Rev. Lett. \textbf{68}, 894 (1992).

\bibitem{karimi10}  E. Karimi, J. Leach, S Slussarenko, B. Piccirillo, L. Marrucci, L. Chen, W. She, S. Franke-Arnold, M. J. Padgett, and E. Santamato, Spin-orbit hybrid entanglement of photons and quantum contextuality, Phys. Rev. A \textbf{82}, 022115 (2010).

\bibitem{fuwa15} M. Fuwa, S. Takeda, M. Zwierz, H. M. Wiseman, and A. Furusawa , Experimental proof of nonlocal wavefunction collapse for a single particle using homodyne measurements. Nat Commun \textbf{6}, 6665 (2015).

\bibitem{george13} R. E. George, L. M. Robledo, O. J. E. Maroney, M. S. Blok, H. Bernien, M. L. Markham, D. J. Twitchen, J. J. L. Morton, G. Andrew D. Briggs, and Ronald Hanson, Opening up three quantum boxes causes classically undetectable wavefunction collapse, PNAS, \textbf{110} (10), 3777-3781 (2013 ).

\bibitem{garrisi} Francesco Garrisi, Micol Previde Massara, Alberto Zambianchi, Matteo Galli, Daniele Bajoni , Alberto Rimini, and Oreste Nicrosini, Experimental test of the collapse time of a delocalized photon state, Sci Rep \textbf{9}, 11897 (2019).


\bibitem{guerreiro12} Thiago Guerreiro, Bruno Sanguinetti, Hugo Zbinden, Nicolas Gisin, and Antoine Suarez, Single-photon space-like antibunching,  Phys. Lett. A \textbf{376}, 2174–2177  (2012) .

\bibitem{brunner14b}  T. Guerreiro, F. Monteiro, A. Martin, J. B. Brask, T. V\'ertesi, B. Korzh, M. Caloz, F. Brussi\'eres, V. B. Verma, A. E. Lita, R. P. Mirin, S. W. Nam, F. Marsilli, M. D. Shaw, N. Gisin, N. Brunner, H. Zbinden, and R. T. Thew, Demonstration of Einstein-Podolsky-Rosen Steering Using Single-Photon Path Entanglement and Displacement-Based Detection, Phys. Rev. Lett. \textbf{117}, 070404 (2016).

\bibitem{brunner20} Private communication.

\bibitem{azzini20}  S. Azzini  S. Mazzucchi  V. Moretti,  D. Pastorello, and L. Pavesi, Single‐Particle Entanglement, Adv. Quantum Technol. , 2000014 (2020). 

\bibitem{dasenbrook16}  D. Dasenbrook, J. Bowles, J. B. Brask, P. P. Hofer, C. Flindt, and N. Brunner, Single-electron entanglement and nonlocality, New Journal of Physics \textbf{18}, 043036 (2016).

\bibitem{boyd15} E. Karimi and R. W. Boyd, Classical Entanglement?, Science \textbf{350}, 1172 (2015).

\bibitem{boyd20} D. Paneru, E. Cohen, R. Fickler, R. W. Boyd and  E. Karimi,  Entanglement: quantum or Classical?, Rep. Prog. Phys. \textbf{83}, 064001 (2020).

\bibitem{konrad19} T. Konrad, and A. Forbes, Quantum mechanics and classical light, Contemporary Physics \textbf{60}, 1-22 (2019).

\bibitem{forbes19}.  A. Forbes, A. Aiello, B. Ndagano, Classically Entangled Light, Progress in Optics \textbf{64}, 99 (2019).

\bibitem{stern21} Stern, O. A way towards the experimental examination of spatial quantisation in a magnetic field. Z Phys D - Atoms, Molecules and Clusters \textbf{10}, 114–116 (1988).

\bibitem{weinert95} F. Weinert, Wrong theory—Right experiment: The significance of the Stern-Gerlach experiments, Studies in History and Phiosophy of Science Part B: Studies in History and Philosophy of Modern Physics 26 75–86 (1995).

\bibitem{are2017a} E. Ben\'itez Rodr\'iguez, L.M. {Ar\'evalo Aguilar}, and E.Piceno Mart\'inez, A full quantum analysis of the Stern-Gerlach experiment using the evolution operator method: Analysing current issues in teaching quantum mechanics, Eur. J. Phys. \textbf{38}, 025403 (2017).

\bibitem{are2017b} E. Ben\'itez Rodr\'iguez, L.M. {Ar\'evalo Aguilar}, and E.Piceno Mart\'inez, Corrigendum: ‘A full quantum analysis of the Stern–Gerlach experiment using the evolution oper- ator method: Analysing current issues in teaching quan- tum mechanics’, Eur. J. Phys. \textbf{38}, 069501 (2017).

\bibitem{paris13} C. Sparaciari and M. G. A. Paris, Canonical naimark extension for generalized measurements involving sets of Pauli quantum observables chosen at random, Phys. Rev. A \textbf{87}, 012106 (2013).

\bibitem{are19} J.A. Mendoza Fierro and L.M. {Ar\'evalo Aguilar}, Stern-Gerlach experiment with arbitrary spin: Temporal evolution and entanglement, Eur. Phys. J. Plus 1\textbf{34}, 82  (2019).

\bibitem{elena2018} A. E. Piceno Mart\'inez, E. Ben\'itez Rodr\'iguez, J. A.
Mendoza Fierro, M. M. M\'endez Otero, and L. M. Ar\'evalo  Aguilar, Quantum Nonlocality and Quantum Correlations in the Stern–Gerlach Experiment, Entropy \textbf{20}, 299 (2018).

\bibitem{margalit19} Y. Margalit1, Z. Zhou, S. Machluf, Y. Japha, S. Moukouri and R. Folman, Analysis of a high-stability Stern–Gerlach spatial fringe interferometer, New J. Phys. \textbf{21} 073040 (2019).

\bibitem{machluf13} Machluf S, Japha Y and Folman R 2013 Coherent Stern–Gerlach momentum splitting on an atom chip Nat. Commun. \textbf{4} 2424 (2013).

\bibitem{wiseman07} H. M. Wiseman, S. J. Jones, and A. C. Doherty, Steering, entanglement, nonlocality, and the Einstein-Podolsky-Rosen Paradox, Phys. Rev. Lett. \textbf{98}, 140402 (2007).

\bibitem{jones07}  S. J. Jones, H. M. Wiseman, and A. C. Doherty, Entanglement, Einstein-Podolsky-Rosen correlations, Bell nonlocality, and steering, Phys. Rev. A \textbf{76}, 052116 (2007).

\bibitem{cavalcanti09}  E. G. Cavalcanti, S. j. Jones, H. M. Wiseman, and M. D. Reid, Experimental criteria for steering and the Einstein-Podolsky-Rosen paradox, Phys. Rev. A \textbf{80}, 032112 (2009).

\bibitem{uola20}  R. Uola, A. C. S. Costa, H. C. Nguyen, and O. G\"une, Quantum steering, Rev. Mod. Phys. \textbf{92}, 015001 (2020).

\bibitem{brunner14}  N. Brunner, D. Cavalcanti, S. Pironio, V. Scarini, and S. Wehner, Bell nonlocality, Rev. Mod. Phys. \textbf{86}, 419 (2014).

\bibitem{cao} H. Cao and Z. Guo, Characterizing Bell nonlocality and EPR steering, Sci. China Phys. Mech. Astron. \textbf{62}, 30311 (2019).

\bibitem{paris20}  M. Frigerio, C. Destri, S. Olivares, and Matteo G. A. Paris, Nonclassical steering with two-mode gaussian states, arXiv:2005.00046 [quant-ph] (2020).

\bibitem{cohen-tannoudji2019} Claude Cohen-Tannoudji, Bernard Diu, and Franck Lalo\"e, QUANTUM MECHANICS Volume I,
Basic Concepts, Tools, and Applications. WILEY-VCH (2020).

\bibitem{hofmann17a} H. F. Hofmann, Quantum interfeence of position and momentum: A particle propagation paradox, Phys. Rev. A \textbf{96}, 020101 (2017).

\bibitem{hofmann17} T. Nii, M. Iinuma, H. F. Hofmann, On the relation between measurement outcomes and physical properties, Quantum Stud.: Math. Found. \textbf{5}, 229 (2018).



\bibitem{li15} Li, Che-Ming and Chen, Yueh-Nan and Lambert, Neill and Chiu, Ching-Yi and Nori, Franco, Certifying single-system steering for quantum-information processing, Phys. Rev. A \textbf{92}, 062310 (2015).

\end{thebibliography}

\end{document}